\documentclass[11pt]{article}

\usepackage[utf8]{inputenc}
\usepackage[margin=1in]{geometry}
\usepackage{amsmath,amssymb,amsthm,mathtools}
\usepackage{bm}
\usepackage{graphicx}
\usepackage{booktabs}
\usepackage{threeparttable}
\usepackage{placeins}
\usepackage{setspace}
\usepackage{authblk}
\usepackage{xcolor}
\usepackage{xurl} 
\usepackage[colorlinks=true,allcolors=blue,
            bookmarksopen=true,bookmarksnumbered=true,
            breaklinks=true]{hyperref}
\usepackage[capitalize,noabbrev]{cleveref}
\usepackage[shortcuts]{extdash}

\usepackage[backend=biber, 
			style=numeric, 
			sorting=none,
			style=numeric, 
			citestyle=numeric-comp,
			giveninits=true,
			maxbibnames=6, 
			minbibnames=3, 
			url=false, 
			doi=true, 
			isbn=false]{biblatex}
\let\cite\supercite
\DeclareSourcemap{ \maps[datatype=bibtex]{ \map{ \step[fieldset=month, null] } } }

\DeclareMathOperator{\E}{E}
\DeclareMathOperator{\Var}{Var}
\DeclareMathOperator{\Cov}{Cov}
\DeclareMathOperator{\Cor}{Cor}

\DeclareMathOperator*{\argmin}{arg\,min}

\newcommand{\trans}{^{\top}}
\newcommand{\dd}{\,\mathrm{d}}
\newcommand{\mathperiod}{\,\text{.}}
\newcommand{\mathcomma}{\,\text{,}}

\newcommand{\semx}{\mathrm{SEM\text{-}X}}
\newcommand{\mathbbm}[1]{\mathbf{#1}}
\newtheorem*{example}{Example}

\crefname{equation}{Equation}{Equations}
\Crefname{equation}{Equation}{Equations}
\creflabelformat{equation}{#2#1#3}

\title{Combining Covariate Adjustment with Information from Secondary Endpoints
to Improve Precision in Randomized Trials}

\author[1]{\small Jack M. Wolf\thanks{Corresponding author: \href{mailto:jack.wolf@pennmedicine.upenn.edu}{jack.wolf@pennmedicine.upenn.edu}}}
\author[2]{\small Joseph S. Koopmeiners}
\author[2]{\small David M. Vock}
\affil[1]{\small Department of Biostatistics, Epidemiology and Informatics,
Perelman School of Medicine, University of Pennsylvania, Philadelphia, PA, USA}
\affil[2]{\small Division of Biostatistics \& Health Data Science,
School of Public Health, University of Minnesota, Minneapolis, MN, USA}

\date{}

\begin{document}
\maketitle

\begin{abstract}
\small
\noindent\textit{Background/Aims}: Adjustment for prognostic baseline covariates can improve precision in randomized trials. Previous work has shown that jointly modeling primary and secondary endpoints can yield additional precision by borrowing information across endpoints. We investigated whether these approaches can be combined to achieve efficiency gains beyond those obtained through covariate adjustment alone.

\noindent\textit{Methods}: We extended a previously proposed one-factor structural equation modeling framework for borrowing information from secondary endpoints to incorporate baseline covariates while retaining the average treatment effect on the primary endpoint as the estimand. To mitigate sensitivity to model misspecification, we combined this estimator with a conventional covariate-adjusted estimator using cross-validated model averaging. We evaluated operating characteristics in simulations and applied the methods to a randomized trial of very low versus normal nicotine content cigarettes.

\noindent\textit{Results}: When the structural equation model was correctly specified, endpoint borrowing improved efficiency beyond conventional covariate adjustment across all simulated settings. Model misspecification could induce bias and undercoverage. Model averaging reduced bias and improved coverage relative to the structural equation model estimator, although coverage remained imperfect under severe misspecification. In the trial application, the model-averaged estimate was 21\% more precise than the unadjusted estimate and 13\% more precise than covariate adjustment alone.

\noindent\textit{Conclusion}: Secondary endpoints can contribute meaningful information about the average treatment effect on a primary endpoint even after baseline covariates have been incorporated. These gains require stronger assumptions than conventional covariate adjustment; model averaging provides a practical compromise between efficiency and robustness.
\end{abstract}


\vfill
\doublespacing

\section{Introduction}

Randomized controlled trials (RCTs) provide the foundation for causal evaluation of interventions, but their cost, duration, and recruitment demands often limit the sample size.
Even when a trial is adequately powered for its final primary analysis, information may remain limited at interim analyses, within clinically important subgroups, or for rare outcomes of scientific interest; in rare-disease trials, recruitment constraints may limit power even for the primary analysis.

These limitations have motivated a broad literature on more efficient analytical methods for RCTs.
Adjustment for baseline covariates is among the most widely applicable approaches.
Early work emphasized the role of covariates in addressing chance imbalance and improving precision\cite{senn_covariate_1989}.
Subsequent theoretical work clarified when covariate-adjusted estimators are consistent, efficient, and robust\cite{yang_efficiency_2001, tsiatis_covariate_2008, freedman_regression_2008, lin_agnostic_2013, wang_analysis_2019}.
More recent work has emphasized alignment between adjustment methods and marginal estimands, model-robust inference, and applicability across randomization schemes\cite{ye_toward_2023, van_lancker_covariate_2024, bannick_covariate_2026}.
Recent regulatory guidance has likewise stressed the need to prespecify covariate-adjusted analyses and to justify the choice of adjustment covariates in confirmatory trials\cite{european_medicines_agency_guideline_2015, food_and_drug_administration_adjusting_2024}.

However, covariate adjustment only uses information that is measured before treatment assignment.
Many trials also collect multiple post-randomization efficacy endpoints in addition to the primary endpoint, such as biomarkers.
These endpoints may characterize the same underlying disease process, clinical construct, or treatment-response pathway and can therefore contain information relevant to the treatment effect on the primary endpoint.
For example, RCTs in tobacco regulatory science may evaluate abstinence from smoking while also measuring biomarkers of nicotine and toxicant exposure.
Similarly, cancer trials may evaluate progression-free survival while also repeatedly measuring circulating tumor DNA as a marker of evolving tumor burden. 

When used appropriately, this information from secondary endpoints may improve power and efficiency.
Global testing procedures combine evidence across several endpoints but generally cannot estimate treatment effects on specific endpoints\cite{obrien_procedures_1984}.
Aggregating information through composite and hierarchical endpoints can likewise improve power, but doing so redefines the estimand and complicates interpretation\cite{freemantle_composite_2003,pocock_win_2012}.
Other recent approaches have used secondary endpoints to facilitate adaptive borrowing from external data sources\cite{wolf_leveraging_2024}.
Together, these approaches highlight the potential value of secondary endpoints, but they do not directly address how to use them to improve precision for the primary-endpoint treatment effect without targeting a different estimand or relying on borrowing across subpopulations or external sources\cite{wolf_commentary_2024}.

We recently proposed a framework that jointly models primary and secondary endpoints using a one-factor structural equation model (SEM) and uses model averaging to mitigate bias due to model misspecification\cite{wolf_jointly_2026}.
This approach targets the treatment effect on the primary endpoint, rather than a composite or global estimand, and can yield substantial efficiency gains versus conventional unadjusted estimators. 
However, the original framework did not incorporate baseline covariates and therefore did not address how endpoint-based information can be combined with baseline covariate adjustment to further improve efficiency.

In this manuscript, we extend the SEM-based endpoint-borrowing framework to incorporate covariate adjustment.
Our objectives are to develop covariate-adjusted estimators that retain the primary-endpoint estimand while leveraging information from both baseline covariates and secondary endpoints, and to assess whether joint modeling of endpoints provides meaningful precision gains beyond those achieved by covariate adjustment alone.
Through simulation studies and an application to a recent RCT evaluating very low nicotine content cigarettes in the presence of alternative nicotine delivery systems, we compare our proposed framework to unadjusted and standard covariate-adjusted approaches.
This work connects two complementary routes toward trial efficiency and provides guidance for analyses in which sample size limitations constrain statistical precision.

\section{Proposed Method}

\subsection{Notation and Target Estimand}

Suppose we observe independent participants indexed by $i=1,\ldots,n$. 
For participant $i$, let $A_i\in\{0,1\}$ denote randomized treatment assignment, $X_i\in\mathbb{R}^q$ denote a vector of baseline covariates, and $Y_i=(Y_{i,1},\ldots,Y_{i,P})\trans$ denote a vector of $P$ post-randomization endpoints, where $Y_{i1}$ is the primary endpoint.
We restrict attention to $P\ge 3$ to ensure identification of the proposed factor model. 
We focus on the average treatment effect on the primary endpoint, $\tau_1\coloneq \E\{Y_{i,1}(1)-Y_{i,1}(0)\}$, where $Y_{i,j}(a)$ denotes the potential outcome on endpoint $j$ that would have been observed if assigned treatment $a$\cite{rubin_estimating_1974}.
Under randomization of treatment and standard causal assumptions, the ATE is identified as the difference between the arm-specific means:
\begin{equation}
    \label{eqn:ate-identification}
    \tau_1= \E(Y_{i,1}\mid A_i=1)-\E(Y_{i,1}\mid A_i=0)\mathperiod
\end{equation}
Applying the law of total expectation, the ATE can also be identified using standardization, in which the conditional mean given $A=a$ and $X$ is averaged over the distribution of $X$ within the population:
\begin{equation}
    \label{eqn:ate-standardization}
    \tau_1 = \E_X\{\E(Y_{i,1}\mid A_i=1,X_i)-\E(Y_{i,1}\mid A_i=0,X_i)\}\mathperiod
\end{equation}

\subsection{Covariate-Adjusted Structural Equation Model Estimator}
We consider an extended SEM that adjusts for baseline covariates.
Let $\eta_i$ denote a latent factor through which treatment affects the endpoints. 
The latent-variable model is 
\[\eta_i \mid A_i \sim N(\gamma A_i,1)\mathperiod\]
Then, conditional on the latent variable and covariates, the endpoint-specific distributions satisfy 
$$Y_{i,p}\mid \eta_i, X_i \sim F_p(\mu_{i,p}, \theta_p), \quad p = 1,\ldots,P\mathcomma$$
with
$$g_p(\mu_{i,p})=\nu_p + K_{p} X_i + \lambda_p \eta_i, \quad \mu_{i,p}=\E(Y_{i,p}\mid \eta_i,X_i)\mathperiod$$
Here, $F_p$ is an endpoint-specific parametric distribution, $g_p$ is a link function, and $\theta_p$ is a nuisance or dispersion parameter when required.
When $F_p$ is an exponential-family distribution, the conditional model for each endpoint is a generalized linear model in which $\eta_i$ enters the linear predictor as a shared random effect.
This formulation accommodates, for example, Gaussian endpoints with identity links, binary endpoints with logit or probit links, and count endpoints with log links. 

Within this working model, treatment shifts the mean of a shared latent factor, and the endpoint distributions depend on treatment only through this latent factor.
Baseline covariates enter the endpoint-specific linear predictors directly through $K_pX_i$.
The endpoints are assumed to be mutually independent conditional on $(\eta_i,X_i)$.
After integrating over $\eta_i$, the shared factor is the sole source of residual dependence among the endpoints conditional on treatment and the baseline covariates.
The parameter $\gamma$ characterizes the treatment-related shift in the latent factor, while $\lambda_p$ determines how strongly $Y_{i,p}$ depends on that factor.
These restrictions enable information sharing across endpoints but also create sensitivity to model misspecification, motivating the model-averaging procedure described below.
Baseline covariates are permitted to have endpoint-specific direct effects and are not required to act through the shared factor.
Hence, unlike the treatment effects, the covariate effects are not constrained to follow a common pattern across endpoints; consequently, the proposed model does not use the secondary endpoints to borrow information directly about the primary endpoint's covariate coefficients.

The joint conditional model can be written as 
$$f(Y_i\mid \eta_i,X_i) = \prod_{p=1}^P f_p(Y_{i,p} \mid \eta_i, X_i; \vartheta_p)\mathcomma \quad \vartheta_p \coloneq (\nu_p, K_p, \lambda_p, \theta_p)\mathcomma$$
and the observed data likelihood contribution is
$$L_i(\vartheta)  = \int_{\mathbb R} \Bigl\{  \prod_{p=1}^P f_p(Y_{i,p} \mid \eta, X_i; \vartheta_p)\Bigr\} \phi(\eta - \gamma A_i) \dd\eta\mathcomma $$
where $\phi$ denotes the standard Gaussian density and $\vartheta \coloneq (\gamma, \vartheta_1\trans,\ldots,\vartheta_P\trans)$ is the full parameter vector.
As the integral may not be evaluable in closed form, the likelihood contribution of each observation can be approximated using Gauss--Hermite quadrature.

In the special case where all $P$ endpoints are modeled as Gaussian with an identity link, the observed data likelihood has the closed form $Y_i \mid A_i,X_i \sim N(\mu_i,\Sigma)$, where
$\mu_i=\nu + KX_i + \gamma\lambda A_i$,
$\Sigma = \operatorname{diag}(\theta_1,\ldots,\theta_P)+\lambda\lambda\trans$,
and $\theta_p$ is the residual variance for $Y_{i,p}$ given $(\eta_i,X_i)$.
Under this model, the endpoint-specific treatment effects are $\tau_p = \gamma\lambda_p$, whereas the conditional covariance between endpoints $j$ and $k$ is $\lambda_j\lambda_k$.
Consequently, for $j\neq k$, $\tau_j\tau_k= \gamma^2\Cov(Y_{i,j},Y_{i,k}\mid A_i,X_i)$.
The treatment-effect vector and the residual covariance structure must therefore be compatible with the same factor loadings. 
For example, the model is incompatible with two positively correlated endpoints having treatment effects in opposite directions, or two negatively correlated endpoints having treatment effects in the same direction.

The ATE is then obtained using standardization (\cref{eqn:ate-standardization}). 
Define the latent-factor-marginalized conditional mean as
$$ m_{p,\semx}(a,x;\vartheta) \coloneq \int_{\mathbb R} g_p^{-1}(\nu_p + K_{p} x + \lambda_p \eta)\phi(\eta-\gamma a)\dd\eta,\quad a\in\{0,1\}\mathperiod$$
This integral can be evaluated in closed form for some distribution-link combinations and otherwise approximated using quadrature.
The model-implied ATE is then
\begin{equation}
    \label{eqn:sem-model-ate}
    \tau_{p,\semx} \coloneq \E_X\{m_{p,\semx}(1,X;\vartheta) - m_{p,\semx}(0,X;\vartheta)\}\mathperiod    
\end{equation}
If the $p$th endpoint is modeled using the identity link, the corresponding ATE reduces to  $\tau_{p,\semx}=\gamma\lambda_p$.

\begin{example}[Binary Primary Endpoint]
    Consider a binary primary endpoint and Gaussian secondary endpoints. 
    We specify 
    $$Y_{i,1}\mid \eta_i,X_i\sim \operatorname{Bernoulli}(p_{i,1}),\quad  g_1(p_{i,1})=\nu_1 + K_1 X_i+\lambda_1\eta_i\mathcomma$$ and, for $p=2,\ldots,P$, 
    $$Y_{i,p}\mid\eta_i,X_i \sim N(\mu_{i,p},\theta_p), \quad \mu_{i,p}=\nu_p + K_p X_i + \lambda_p\eta_i\mathperiod$$
    If $g_1=\Phi^{-1}$ is the probit link, the latent-factor integration has a closed form:
    $$m_{1,\semx}(a,x;\vartheta) = \Phi\left(\frac{\nu_1+K_1 x + \lambda_1\gamma a}{\sqrt{1+\lambda_1^2}} \right)\mathperiod$$
    Alternatively, if $g_1$ is the logit link,
    $m_{1,\semx}(a,x;\vartheta) = \int_{\mathbb R} \operatorname{expit}\{\nu_1 + K_1 x + \lambda_1\eta\}\phi(\eta-\gamma a)\dd \eta$,
    where $\operatorname{expit}(x)=1/(1+e^{-x})$.
    In either case, the marginal risk difference is obtained by standardizing $m_{1,\semx}(a,x;\vartheta)$ via \cref{eqn:sem-model-ate}.
\end{example}

We estimate the ATE using the plug-in principle. 
First, we estimate the SEM parameters via maximum likelihood and obtain $\widehat\vartheta$.
From the fitted model, we estimate the conditional mean of the primary endpoint given treatment $a$ and covariates $x$: $m_{1,\semx}(a,x;\widehat\vartheta)$. 
Then, we average these estimated conditional means over the empirical distribution of the baseline covariates: 
\begin{equation}
    \label{eqn:sem-estimator}
    \widehat\tau_{1,\semx}\coloneq\frac1n \sum_{i=1}^n \{ m_{1,\semx}(1,X_i;\widehat\vartheta) - m_{1,\semx}(0,X_i;\widehat\vartheta)\}\mathperiod    
\end{equation}
If the primary endpoint is modeled using the identity link, the estimator reduces to $\widehat\tau_{1,\semx}=\widehat\gamma\widehat\lambda_1$.

\subsection{Model Averaging}

To balance the potential efficiency gains of the SEM estimator against potential bias from model misspecification, we perform data-adaptive model averaging over a library of candidate estimators. 
We consider a library that contains the proposed covariate-adjusted SEM estimator as well as conventional estimators that remain consistent under weaker assumptions but may be less efficient.
Let $m=1,\ldots,M$ index the candidate primary-endpoint ATE estimators, denoted $\widehat\tau_{1,m}$.
For example, $\widehat\tau_{1,1}=\widehat\tau_{1,\semx}$ may be the covariate-adjusted SEM estimator and $\widehat\tau_{1,2}$ may be a conventional covariate-adjusted estimator, such as ANCOVA for a continuous primary endpoint. 
For weights $\omega\coloneq (\omega_1,\ldots,\omega_M)\trans$ in the unit simplex
$\Delta_M \coloneq \{\omega : \omega_m \geq 0, \sum_{m=1}^M \omega_m=1\}$, 
the model-averaged ATE estimator is
\[\widehat{\tau}_{1,\mathrm{MA}}(\omega) \coloneq \sum_{m=1}^M \omega_m \widehat\tau_{1,m}\mathperiod\]

We select weights using repeated cross-validation targeted to the primary-endpoint ATE, reducing sensitivity to any single fold partition.
Let $R\ge1$ denote the number of cross-validation schedules. For each $r=1,\ldots,R$, independently construct a treatment-stratified cross-validation schedule that partitions the sample into $J$ disjoint and approximately equally sized folds, denoted $V_{r,1},\ldots,V_{r,J}$.
Let $T_{r,j}$ denote the corresponding training sample containing all observations not in $V_{r,j}$, and let $n_{r,j}\coloneq \lvert V_{r,j}\rvert$ denote the size of the validation fold.
For each schedule $r$ and fold $j$, apply each candidate estimator to $T_{r,j}$ and denote the resulting primary-endpoint ATE estimate by $\widehat\tau_{1,m}^{(-r,j)}$.
We then calculate a reference estimate $\widehat\tau_{1,\mathrm{ref}}^{(r,j)}$ using only the observations in $V_{r,j}$.
The reference estimator should target $\tau_1$ and be consistent under trial randomization without relying on the endpoint-borrowing restrictions imposed by the SEM.
Possible choices include the unadjusted difference in arm-specific means, a conventional covariate-adjusted estimator such as ANCOVA, or another design-consistent estimator appropriate for the primary-endpoint type.

We select the model-averaging weights by pooling the ATE-targeted losses across the $R$ cross-validation schedules:
\begin{equation}
    \label{eqn:selected-weights}
    \widehat\omega \coloneq \argmin_{\omega\in \Delta_M}  \sum_{r=1}^R \sum_{j=1}^J n_{r,j} \Bigl( \widehat\tau_{1,\mathrm{ref}}^{(r,j)} - \sum_{m=1}^M \omega_m \widehat\tau_{1,m}^{(-r,j)} \Bigr)^2\mathperiod
\end{equation}
After selecting $\widehat{\omega}$, we apply each candidate estimator to the full sample to obtain $\widehat\tau_{1,m}$.
The final model-averaged estimator is 
\begin{equation}
    \label{eqn:ma-estimator}
    \widehat{\tau}_{1,\mathrm{MA}}\coloneq \widehat\tau_{1,\mathrm{MA}}(\widehat \omega) = \sum_{m=1}^M \widehat \omega_m \widehat\tau_{1,m} \mathperiod
\end{equation}

We estimate the sampling variance of the model-averaged estimator using the nonparametric bootstrap, repeating the complete weight-selection and estimation procedure within each bootstrap resample.
We construct Gaussian-approximation confidence intervals using the bootstrap standard error.

\section{Simulation Studies}

\subsection{Simulation Design}

We considered three simulation experiments to assess the performance of the proposed estimator, mirroring the design of Wolf~et~al.~(2026), while adding prognostic baseline covariates. 
Simulation~1 assessed the potential for efficiency and power gains under correctly specified models with varying baseline prognostic information and residual secondary-endpoint information.
Simulation~2 considered performance in the presence of SEM misspecification under both global alternative (2a) and primary-null (2b) hypotheses for three Gaussian endpoints and a global alternative hypothesis (2c) for a binary primary endpoint.
Finally, Simulation~3 evaluated performance under the global null hypothesis of no effect on any endpoint.
In each scenario, $n=250$ participants were randomized with equal probability to two treatment groups.
Within each simulation study, we manipulated the covariate-endpoint and endpoint-endpoint relationships while holding the arm-specific endpoint means and variances fixed.
The data-generating processes are described in detail in the Supplemental Materials and summarized in \cref{tab:simulation-design}.

We characterized two complementary sources of information about the primary endpoint: prognostic information from baseline covariates and residual information from the secondary endpoints.
For either treatment arm $a$, we quantified baseline prognostic information using 
$$R_{1\mid X}^2\coloneq \frac{\Var\{\E(Y_{i,1}\mid A_i=a,X_i)\mid A_i\}}{\Var(Y_{i,1}\mid A_i=a)}\mathcomma$$ the proportion of primary-endpoint variation explained by baseline covariates.
To isolate the additional information available from the secondary endpoints after covariate adjustment, define the endpoint-specific residuals as $\epsilon_{i,p}\coloneq Y_{i,p}-\E(Y_{i,p}\mid A_i,X_i)$ and let $\epsilon_{i,-1}\coloneq (\epsilon_{i,2},\ldots,\epsilon_{i,P})\trans$ collect the secondary-endpoint residuals.
We then quantified the residual secondary-endpoint information using
$$R_{1\mid -1, X}^2\coloneq \frac{\Var\{\E(\epsilon_{i,1}\mid A_i=a,\epsilon_{i,-1}) \mid A_i=a\}}{\Var(\epsilon_{i,1}\mid A_i=a)}\mathcomma$$
the proportion of primary-endpoint residual variation explained by the secondary endpoint residuals after adjustment for treatment and baseline covariates.  
All simulations manipulated $R_{1\mid -1, X}^2$ along different covariance paths. 
In Simulation 1, we moved the covariance model along a path of covariance matrices compatible with the mean structure implied by the average treatment effect across all endpoints.
In Simulation 2, the primary-secondary covariances were varied while holding the secondary-secondary covariance fixed, thereby inducing model misspecification away from a single compatible reference value. 
For Simulation 2c, the binary primary endpoint was generated by thresholding an underlying Gaussian response under a probit model. Residual secondary-endpoint information was defined using the residual covariance of this underlying response and the continuous secondary endpoints.
Simulation 3 directly varied only one primary-secondary covariance under the global null; this path explored covariance structures both compatible and incompatible with the one-factor SEM.

\begin{table}[tbp]
\centering
\caption{Summary of the simulation design.}
\label{tab:simulation-design}
\small
\begin{threeparttable}
    \begin{tabular}{
        @{}
        p{0.09\textwidth}
        p{0.25\textwidth}
        p{0.18\textwidth}
        p{0.37\textwidth}
        @{}
    }
    \toprule
    Simulation &
    Endpoints and Treatment Effects &
    Feature Varied &
    Purpose and Interpretation \\
    \midrule
    
    1 &
    Three Gaussian endpoints; treatment effects nonzero for all endpoints &
    $R^2_{1\mid X}$ and $R^2_{1\mid-1,X}$ &
    Efficiency and power as baseline prognostic information and residual secondary-endpoint information increase. 
    The covariance and treatment-effect structures are compatible with the adjusted structural equation model throughout. \\
    \addlinespace
    
    2a &
    Three Gaussian endpoints; treatment effects nonzero for all endpoints &
    $R^2_{1\mid-1,X}$ &
    Sensitivity to model misspecification under the alternative. 
    The covariance and treatment-effect structures are compatible at $R^2_{1\mid-1,X}=0.30$. \\
    \addlinespace
    
    2b &
    Three Gaussian endpoints; null primary effect and nonzero secondary-endpoint effects &
    $R^2_{1\mid-1,X}$ &
    Sensitivity to model misspecification with a null effect for the primary endpoint and nonzero effects for the secondary endpoints.
    The covariance and treatment-effect structures are compatible at $R^2_{1\mid-1,X}=0$. \\
    \addlinespace
    
    2c &
    Binary primary endpoint and two Gaussian secondary endpoints; treatment effects nonzero for all endpoints &
    $R^{2,*}_{1\mid-1,X}$ &
    Sensitivity to model misspecification with a binary primary endpoint. 
    The covariance and treatment-effect structures are compatible at $R^{2,*}_{1\mid-1,X}=0.30$. \\
    \addlinespace
    
    3 &
    Three Gaussian endpoints; global treatment null &
    $\rho_{12}$ &
    Sensitivity to model misspecification under the global null hypothesis. 
    Within the simulated range, the covariance and treatment-effect structures are compatible for approximately $0.237 \le\rho_{12}\le 0.767$. \\
    \bottomrule
    \end{tabular}
    \begin{tablenotes}[para,flushleft]
    \footnotesize
    \item[]
    $R^2_{1\mid X}$ is the proportion of primary-endpoint variance explained by
    baseline covariates conditional on treatment.
    $R^2_{1\mid-1,X}$ is the proportion of residual primary-endpoint variance explained by the secondary-endpoint residuals after adjustment for treatment and baseline covariates.
    The asterisk denotes that the quantity is defined on the latent Gaussian scale for the binary primary endpoint.
    $\rho_{12}\coloneq \Cor(Y_{i1},Y_{i,2}\mid A_i,X_i)$ is the conditional correlation between $Y_{i1}$ and $Y_{i2}$.
    Exact data-generating models, parameter grids, and covariance constructions are provided in the Supplemental Materials.
    \end{tablenotes}
\end{threeparttable}
\end{table}

We compared four estimators of the primary-endpoint ATE: (1) the unadjusted difference in arm-specific sample means, $\widehat\tau_{1,\mathrm{DM}}$; (2) a primary-only covariate-adjusted estimator, $\widehat\tau_{1,\mathrm{Adj}}$; (3) the proposed covariate-adjusted SEM estimator, $\widehat\tau_{1,\semx}$; and (4) the model-averaged estimator $\widehat\tau_{1,\mathrm{MA}}$.
For continuous primary endpoints, we obtained the adjusted estimator using ANCOVA with covariate main effects. 
For binary primary endpoints, we obtained the adjusted estimator by fitting a correctly specified probit regression model and standardizing over the empirical covariate distribution.
The model-averaging library included $\widehat\tau_{1, \mathrm{Adj}}$ and $\widehat\tau_{1,\semx}$; we selected weights using five-fold cross-validation aggregated over two cross-validation schedules.
For the unadjusted and ANCOVA estimators, confidence intervals used robust standard errors; standardized probit inference used the model-based delta method. 
For $\widehat\tau_{1,\semx}$ and $\widehat\tau_{1,\mathrm{MA}}$, standard errors were estimated using the nonparametric bootstrap with 200 replicates.

We generated and analyzed 1000 Monte Carlo datasets per scenario.
Within each scenario, we summarized each method's empirical bias, variance, mean squared error, and coverage for $\tau_1$, as well as the rejection rate of the null hypothesis that $\tau_1=0$.

\subsection{Simulation Results}

\begin{figure}
    \centering
    \includegraphics[width=\linewidth]{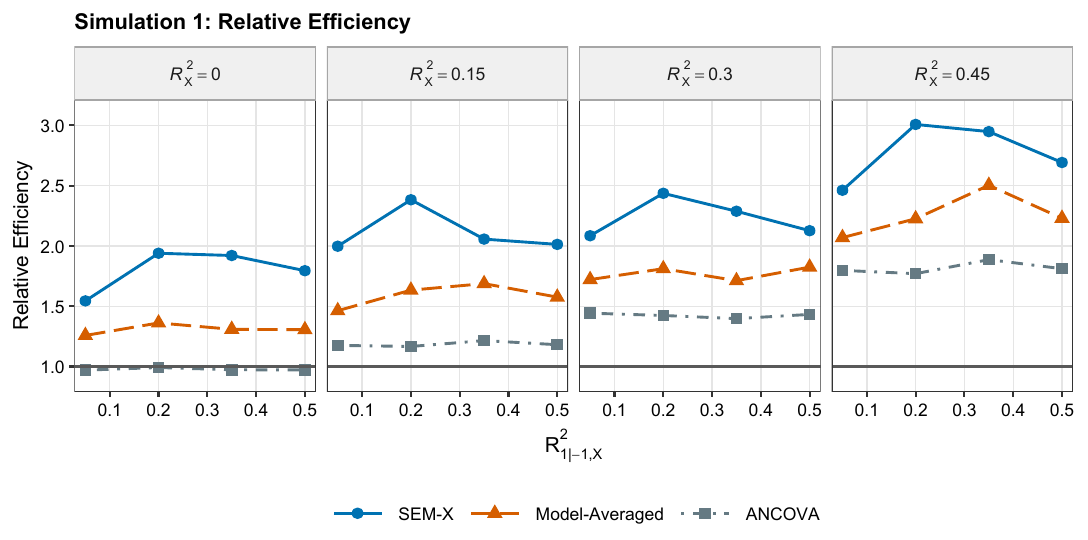}
    \caption{Relative efficiency of $\widehat\tau_{1,\semx}$, $\widehat\tau_{1,\mathrm{MA}}$, and $\widehat\tau_{1,\mathrm{Adj}}$ versus $\widehat\tau_{1,\mathrm{DM}}$ under correct model specification. }
    \label{fig:sim1}
\end{figure}

In Simulation~1, the SEM working model was correctly specified.
Consistent with the existing literature, the efficiency of $\widehat\tau_{1,\mathrm{Adj}}$ relative to $\widehat\tau_{1,\mathrm{DM}}$ increased with the prognostic information in the baseline covariates (\cref{fig:sim1}).
The covariate-adjusted SEM provided additional efficiency gains in every scenario, including when the residual secondary-endpoint information was weak. 
The model-averaged estimator was generally more efficient than $\widehat\tau_{1,\mathrm{Adj}}$ but less efficient than $\widehat\tau_{1,\semx}$.

\begin{figure}
    \centering
    \includegraphics[width=\linewidth]{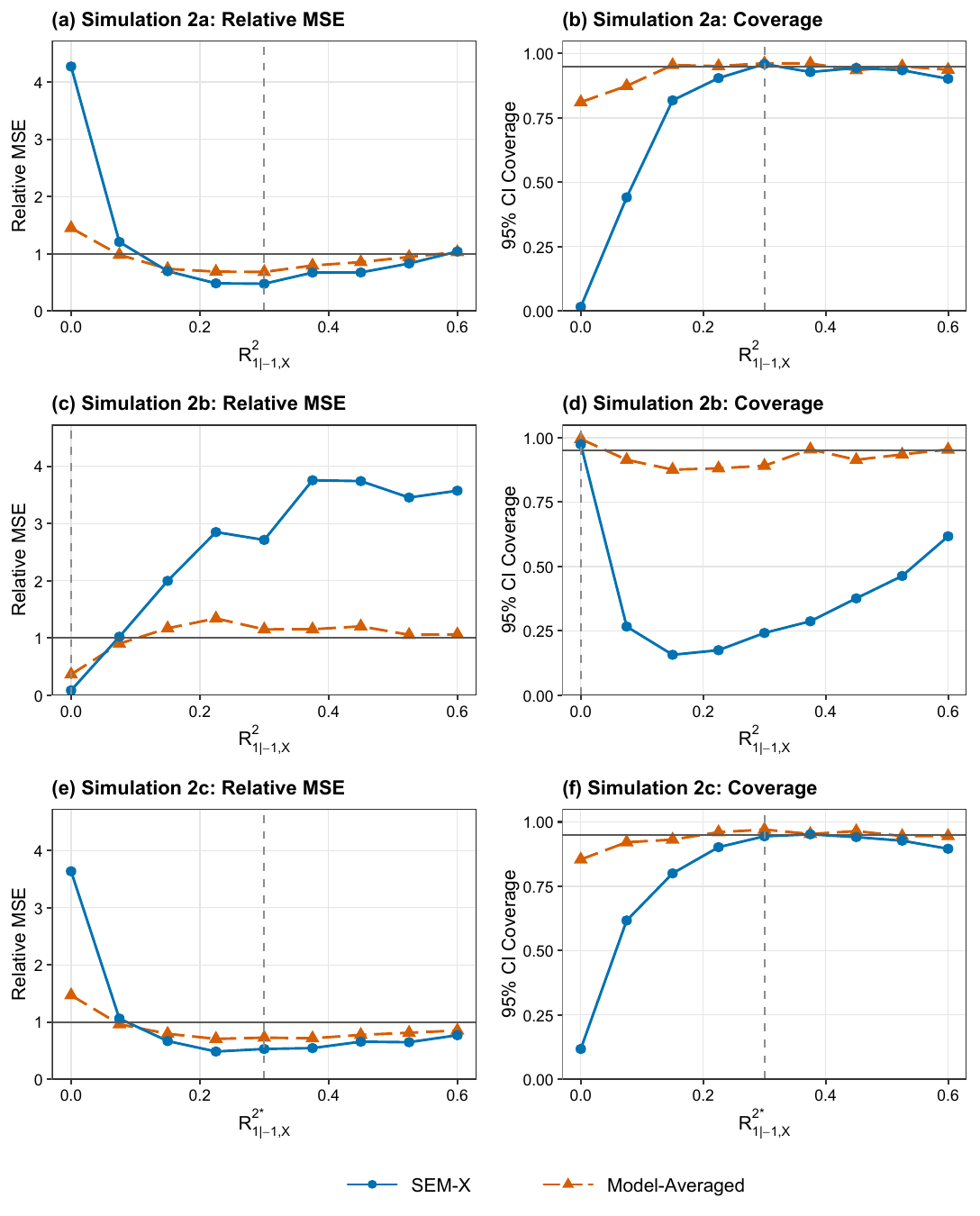}
    \caption{Relative mean squared error (MSE) versus $\widehat\tau_{1,\mathrm{Adj}}$ and coverage of 95\% confidence intervals for $\widehat\tau_{1,\semx}$ and $\widehat\tau_{1,\mathrm{MA}}$ under model misspecification. Dashed vertical reference lines denote the covariance compatible with the structural equation model.}
    \label{fig:sim2}
\end{figure}

Simulation~2 demonstrated the primary motivation for model averaging.
At the setting-specific reference values of $R_{1\mid-1,X}^2$ where the SEM was correctly specified, $\widehat\tau_{1,\semx}$ produced substantial MSE reductions versus $\widehat\tau_{1,\mathrm{Adj}}$ with approximately nominal coverage (\cref{fig:sim2}).
Away from these values, however, $\widehat\tau_{1,\semx}$ became biased, with increased MSE and undercoverage; this sensitivity was greatest in Simulation~2b where a null effect on the primary endpoint coexisted with nonzero effects on correlated secondary endpoints. 
Model averaging adapted by shifting weight toward $\widehat\tau_{1,\mathrm{Adj}}$, substantially attenuating the excess MSE and coverage failures attributable to $\widehat\tau_{1,\semx}$ under model misspecification.
It therefore offered a compromise between efficiency near the working model and robustness away from it, although its coverage was not uniformly nominal.
The binary primary-endpoint setting exhibited the same qualitative bias-variance tradeoff as in the Gaussian setting (Simulation~2a).

Under the global null, setting $\gamma=0$ matches the mean structure and removes the restriction tying the factor loadings to the treatment effects; model compatibility is then determined by whether the residual covariance admits a one-factor representation.
Hence, a wide range of endpoint covariances were compatible with the one-factor SEM in Simulation~3. 
Both $\widehat\tau_{1,\semx}$ and $\widehat\tau_{1,\mathrm{MA}}$ remained approximately unbiased and were more efficient than $\widehat\tau_{1,\mathrm{Adj}}$, even when the covariance structure was incompatible with the SEM. 
Inference for $\widehat\tau_{1,\mathrm{MA}}$ was conservative, whereas inference based on $\widehat\tau_{1,\semx}$ approximately attained the nominal type~I error rate.
Detailed results are provided in the Supplemental Materials.

\section{Application to Tobacco Regulatory Science}

To reduce smoking-related disease and death, the U.S. Food and Drug Administration has proposed a product standard that would limit nicotine in cigarettes\cite{food_and_drug_administration_fda_2025}.
Whether access to very low nicotine content (VLNC) cigarettes promotes abstinence when alternative nicotine products such as e-cigarettes are available is directly relevant to the potential public health impact of such a standard. 

We analyzed data from a recent randomized trial of VLNC versus normal nicotine content (NNC) cigarettes in the presence of alternative nicotine delivery systems\cite{hatsukami_reduced_2024}.
In the trial, $n=438$ participants participated in a virtual marketplace in which they could exchange points for alternative nicotine products as well as either VLNC or NNC cigarettes, depending on their randomization status.
The study was powered to detect an effect on average cigarettes smoked per day 12 weeks following randomization.
Although abstinence was assessed as a secondary endpoint, the study was not powered for this comparison.
We therefore reanalyzed these data to estimate the effect of access to the VLNC marketplace on 12-week point-prevalence abstinence, using both baseline covariates and secondary biomarkers to improve precision.

In addition to measuring point-prevalence abstinence, the trial collected several biomarkers of interest.
These included 
2\=/cyanoethylmercapturic acid (CEMA), a urinary biomarker of recent exposure to combusted cigarette smoke, 
breath carbon monoxide (CO), a biomarker of very recent exposure to combustible smoke, 4\=/(methylnitrosamino)-1-(3-pyridyl)-1-butanol and its glucuronides (NNAL), a biomarker of longer-term exposure to tobacco-specific carcinogens,
and total nicotine equivalents (TNE), a biomarker of total nicotine intake across combustible and noncombustible products.
We hypothesized that the intervention affected these biomarkers and abstinence through a common latent dimension of cigarette-smoking and nicotine-product use behavior and that a one-factor structural equation model may therefore provide a useful approximation to their joint distribution.

Data were available for four biomarkers; however, only two are required to identify the SEM.
A model containing all four biomarkers might provide greater efficiency but would impose restrictions across all five endpoints, increasing sensitivity to misspecification.
We instead included SEM estimators for all six biomarker pairs and the covariate-adjusted probit estimator in the model-averaging library, following the strategy explored previously\cite{wolf_jointly_2026}.
This library allows the analysis to borrow information under several plausible working models without relying exclusively on any single SEM specification.

We estimated the treatment effect on abstinence using the unadjusted risk difference, a standardized covariate-adjusted probit model, each of the six covariate-adjusted SEMs, and model averaging over the probit and SEM estimators.
All adjusted models included age, menthol-cigarette use, baseline cigarettes per day, and baseline Fagerstr\"om Test for Nicotine Dependence (FTND) scores\cite{heatherton_fagerstrom_1991}.
We addressed missing data using multiple imputation by chained equations, with ten imputed datasets and biomarker values from earlier time points used as auxiliary variables.
We selected model-averaging weights separately within each imputed dataset using the same cross-validation scheme as in the simulations. 
We estimated within-imputation variances for the SEM and model-averaged estimators using 100 bootstrap samples. 
We pooled estimates across the imputed datasets, accounting for within- and between-imputation variability\cite{donald_b_rubin_multiple_1987}.

\begin{figure}
    \centering
    \includegraphics[width=\linewidth]{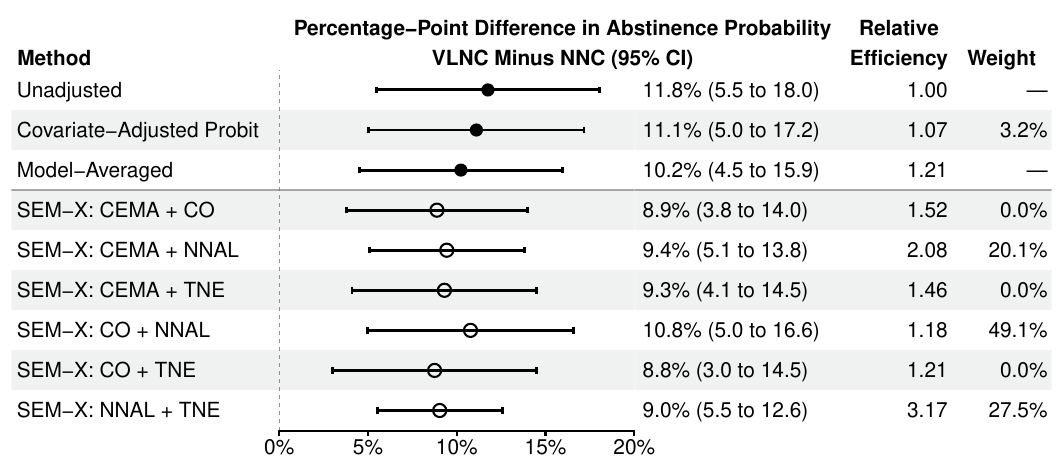}
    \caption{Estimated effect of assignment to the very low nicotine content (VLNC) versus normal nicotine content (NNC) marketplace on 12-week point-prevalence abstinence.
    Effects are presented on the additive scale and represent the percentage-point difference in abstinence probability (VLNC minus NNC).
    Filled points denote the primary comparison estimators, and open points denote the individual covariate-adjusted structural equation model estimators included in the model-averaging library.
    The weight column indicates the corresponding weight in the model-averaged estimator for each method included in the library, averaged across imputations.}
    \label{fig:marketplace-results}
\end{figure}

All methods estimated increases of 8.8--11.8 percentage points in the probability of abstinence under the VLNC marketplace (\cref{fig:marketplace-results}).
The adjusted probit estimate was similar to the unadjusted estimate and was 7\% more efficient.
The individual SEM estimates were smaller and 18--217\% more efficient than the unadjusted estimator, illustrating the potential precision gains but also the dependence of the point estimates on the working-model restrictions.
The model-averaged estimate fell between the adjusted probit and SEM estimates, and placed nonzero weight on the probit estimator and all three SEM estimates that included NNAL in their biomarker pairs: CEMA--NNAL, CO--NNAL, and NNAL--TNE.
The corresponding estimate of a 10.2 (95\% CI: 4.5--15.9) percentage point increase in the probability of abstinence is 21\% more precise than the unadjusted estimate, and 13\% more precise than covariate adjustment alone. 

\section{Discussion}

We developed a SEM-based endpoint-borrowing framework to combine two complementary approaches to improving precision in randomized trials: adjustment for baseline covariates and joint modeling of post-randomization endpoints, such as biomarkers.
In our simulations, the covariate-adjusted SEM provided efficiency gains beyond conventional covariate adjustment when the working model was correctly specified.
These gains persisted across the considered levels of residual secondary-endpoint information, including settings in which that information was weak. 
In an application to tobacco regulatory science, model averaging over a conventional adjusted estimator and several biomarker-based SEM estimators improved precision by 13\% beyond covariate adjustment alone. 
This finding provides an empirical example in which secondary endpoints contributed meaningful additional information after prognostic baseline covariates had been incorporated.
By allowing endpoint-specific distributions and link functions, the proposed formulation also broadens our previous framework beyond the Gaussian and probit models considered there.

These potential efficiency gains rely on the joint endpoint working model.
In our previous unadjusted framework, the endpoints were assumed to be conditionally independent given the shared latent factor, requiring that factor to account for all within-arm dependence among the endpoints.
The inclusion of baseline covariates in the model---beyond improving efficiency---softens this restriction: endpoints are now assumed to be conditionally independent given both the latent factor and the covariates, allowing baseline characteristics to account for some of their association.
Nevertheless, all residual dependence remaining after covariate adjustment must still be represented by a single shared factor, which remains a strong working assumption.

Our simulations examined several consequential departures from this assumption but cannot establish robustness to all forms of misspecification.
In particular, the conditional covariance matrix of the endpoint vector may vary across treatment arms or covariate strata. 
Such multivariate heteroskedasticity may involve changes in marginal variances, pairwise correlations, or both, and could affect estimation and inference. 
Misspecification of the covariate effects, such as omitted nonlinearities or interactions, may also have different consequences for the joint SEM than for conventional covariate adjustment.

Together, these developments and findings highlight a central tradeoff between efficiency and robustness. 
Conventional covariate-adjusted estimators can remain consistent under treatment randomization even when the outcome regression model is misspecified, whereas violations of the joint endpoint restrictions can bias the SEM estimator. 
Model averaging reduced the resulting bias and coverage problems in many of our simulations but did not eliminate them under severe forms of misspecification.
In the application, the precision and, to a lesser extent, the point estimates of the individual SEM estimators varied across biomarker pairs, illustrating that the magnitude of the efficiency gain depends on the selected working model.
Averaging across several SEM specifications and the conventional adjusted estimator provides a practical way to avoid relying exclusively on any single joint model, although it does not guarantee protection from misspecification.
Unlike conventional covariate adjustment, borrowing from biomarkers or other secondary endpoints exchanges stronger modeling assumptions for potential efficiency gains; its use should therefore require correspondingly stronger scientific and statistical justification.

\section*{Acknowledgments}

The authors thank their collaborator, Dr. Dorothy K. Hatsukami, for providing access to the data used in this manuscript.

\section*{Declaration of Conflicting Interests}
The authors declare that there is no conflict of interest.

\section*{Funding}

This study was funded by the National Institute on Drug Abuse (Award Numbers R01DA046320 and U54DA031659). 
The content is solely the responsibility of the authors and does not necessarily represent the official views of the National Institutes of Health and the Food and Drug Administration Center for Tobacco Products.

\printbibliography

\clearpage
\begingroup
\renewcommand{\thesection}{S\arabic{section}}
\renewcommand{\thesubsection}{\thesection.\arabic{subsection}}
\renewcommand{\thesubsubsection}{\thesubsection.\arabic{subsubsection}}
\renewcommand{\theequation}{S\arabic{equation}}
\renewcommand{\thefigure}{S\arabic{figure}}
\renewcommand{\thetable}{S\arabic{table}}
\renewcommand{\theHsection}{supplement.\arabic{section}}
\renewcommand{\theHsubsection}{supplement.\arabic{section}.\arabic{subsection}}
\renewcommand{\theHequation}{supplement.\arabic{equation}}
\renewcommand{\theHfigure}{supplement.\arabic{figure}}
\renewcommand{\theHtable}{supplement.\arabic{table}}
\setcounter{section}{0}
\setcounter{subsection}{0}
\setcounter{equation}{0}
\setcounter{figure}{0}
\setcounter{table}{0}

\let\manuscriptsection\section
\renewcommand{\section}{\FloatBarrier\manuscriptsection}

{\noindent \LARGE \textbf{Supplemental Materials}}
\section{Simulation Study Data Generating Processes}
\label{sec:sim-dgp}

\subsection{Common Data-Generating Model}

Each simulation included one primary endpoint and two secondary endpoints.
For participant $i$, let $X_i=(X_{i1},X_{i2},X_{i3})\trans$ denote the baseline covariates and let $A_i$ denote treatment assignment.
We independently generated
$$X_i\sim N_3(0,I_3), \qquad A_i\sim\operatorname{Bernoulli}(0.5)\mathperiod$$
For a latent response vector $Y_i^*=(Y_{i1}^*,Y_{i2}^*,Y_{i3}^*)\trans$, the common generating model was
\begin{equation}
    \label{eqn:common-dgp}
    Y_i^*=\alpha+KX_i+\tau A_i+\epsilon_i, \qquad \epsilon_i\sim N_3(0,\Sigma_\epsilon)\mathperiod
\end{equation}
Here, $\alpha$ is a vector of intercepts, $K$ is a $3\times3$ matrix of covariate coefficients, $\tau$ is a vector of treatment coefficients, and $\Sigma_\epsilon$ is the residual endpoint covariance matrix.
For Simulations 1, 2a, 2b, and 3, all endpoints were Gaussian and $Y_i=Y_i^*$.
For Simulation 2c, the primary endpoint was binary and generated as $Y_{i1}=\mathbbm{1}(Y_{i1}^*>0)$, while $Y_{i2}=Y_{i2}^*$ and $Y_{i3}=Y_{i3}^*$ remained Gaussian.

We partitioned the residual covariance as
\begin{equation}
    \label{eqn:residual-partition}
    \Sigma_\epsilon=\begin{pmatrix}\sigma_1^2&q\trans\\q&S\end{pmatrix},
\end{equation}
where $q=\Cov(\epsilon_{i1},\epsilon_{i,-1})$ contains the residual covariances between the primary and secondary endpoints and $S=\Var(\epsilon_{i,-1})$ is the residual covariance matrix of the secondary endpoints.
For the Gaussian designs, the residual secondary-endpoint information was
$$R^2_{1\mid-1,X}=\frac{q\trans S^{-1}q}{\sigma_1^2}\mathperiod$$
For Simulation 2c, the analogous quantity $R^{2,*}_{1\mid-1,X}$ was defined using the latent Gaussian residuals.

\subsection{Covariate Coefficient Path}

We began with the coefficient pattern
\begin{equation}
    \label{eqn:k0}
    K_0=\begin{pmatrix}1&0.5&-0.5\\0.8&0.4&-0.4\\0.6&0.3&-0.3\end{pmatrix}\mathperiod
\end{equation}
Let $K_{0,1}$ denote the first row of $K_0$ and let $\Sigma_X=\Var(X_i)=I_3$.
For a target value $r_X$, we set
\begin{equation}
    \label{eqn:k-scaling}
    K(r_X)=\sqrt{\frac{r_X}{K_{0,1}\Sigma_XK_{0,1}\trans}}K_0\mathperiod
\end{equation}
For the Gaussian designs, this construction gives
$$R^2_{1\mid X}=\frac{K_1\Sigma_XK_1\trans}{K_1\Sigma_XK_1\trans+\sigma_1^2}=K_1\Sigma_XK_1\trans=r_X,$$
because the total within-arm variance of the primary endpoint was fixed at one by setting $\sigma_1^2=1-r_X$.

Simulation 1 used
$$r_X\in\{0,0.15,0.30,0.45\}\mathperiod$$
Simulations 2a, 2b, and 3 fixed $r_X=0.30$.
Simulation 2c used the same coefficient matrix but fixed the primary latent residual variance at one for probit identification; its latent-scale proportion of variance attributable to the covariates was therefore $0.30/(1+0.30)\approx0.23$.
The common coefficient matrix used outside Simulation 1 was
\begin{equation}
    \label{eqn:k30}
    K(0.30)=\begin{pmatrix}0.45&0.22&-0.22\\0.36&0.18&-0.18\\0.27&0.13&-0.13\end{pmatrix}\mathperiod
\end{equation}

\subsection{Simulation 1: Compatible Covariance Path}

Simulation 1 used three Gaussian endpoints with $\alpha=(0,0,0)\trans$ and $\tau=(0.25,0.35,0.30)\trans$.
The treatment and residual covariance structures were constructed to remain compatible with the one-factor SEM throughout the simulation grid.
Specifically, we wrote
\begin{equation}
    \label{eqn:compatible-factor}
    \Sigma_\epsilon=\lambda\lambda\trans+\operatorname{diag}(\theta), \qquad \lambda=a u, \qquad u=\frac{\tau}{\lVert\tau\rVert}, \qquad \gamma=\frac{\lVert\tau\rVert}{a}\mathperiod
\end{equation}
This construction ensures $\tau=\gamma\lambda$, as required by the generating one-factor mean structure.

For a selected value of $r_X$, let $d_p=1-K_p\Sigma_XK_p\trans$ denote the residual variance required to retain unit within-arm variance for endpoint $p$.
For a candidate value of $a$, we set $\theta_p=d_p-a^2u_p^2$ and selected $a$ numerically so that the resulting covariance matrix satisfied
$$\frac{q(a)\trans S(a)^{-1}q(a)}{\sigma_1^2(a)}=r_\epsilon\mathperiod$$
The residual-information grid was
$$r_\epsilon\in\{0.05,0.20,0.35,0.50\}\mathperiod$$
Crossing this grid with the four values of $r_X$ produced 16 scenarios.
Because the loading direction remained proportional to $\tau$ while its magnitude was recalibrated, all three cross-endpoint residual covariances generally changed along this path, and the data-generating covariance remained compatible with the SEM in every scenario.

\subsection{Simulations 2a and 2b: Gaussian Covariance-Perturbation Path}

Simulations 2a and 2b used $K(0.30)$ from \cref{eqn:k30} and began with the compatible Gaussian reference covariance
\begin{equation}
    \label{eqn:gaussian-reference}
    \Sigma_{\epsilon,0}=\begin{pmatrix}0.70&0.39&0.34\\0.39&0.81&0.47\\0.34&0.47&0.89\end{pmatrix}=\begin{pmatrix}\sigma_{1,0}^2&q_0\trans\\q_0&S_0\end{pmatrix}\mathperiod
\end{equation}
This reference covariance was calibrated under $\tau=(0.25,0.35,0.30)\trans$ to attain $R^2_{1\mid-1,X}=0.30$.

For a selected grid value $r$, we varied only the primary--secondary residual covariance vector:
\begin{equation}
    \label{eqn:q-perturbation}
    q(r)=\sqrt{r/0.30}\,q_0\mathperiod
\end{equation}
We then set
\begin{equation}
    \label{eqn:gaussian-perturbation}
    \Sigma_\epsilon(r)=\begin{pmatrix}\sigma_{1,0}^2&q(r)\trans\\q(r)&S_0\end{pmatrix}\mathperiod
\end{equation}
The primary residual variance, both secondary residual variances, and the secondary--secondary residual covariance were therefore held fixed.
Moreover,
$$\frac{q(r)\trans S_0^{-1}q(r)}{\sigma_{1,0}^2}=\frac{r}{0.30}\frac{q_0\trans S_0^{-1}q_0}{\sigma_{1,0}^2}=r,$$
so the grid coordinate was exactly $R^2_{1\mid-1,X}$.
We used
$$r\in\{0,0.075,0.150,0.225,0.300,0.375,0.450,0.525,0.600\}\mathperiod$$

Simulation 2a used $\alpha=(0,0,0)\trans$ and $\tau_{\mathrm{2a}}=(0.25,0.35,0.30)\trans$.
The covariance and treatment-effect structures were compatible at $r=0.30$.
Simulation 2b instead used $\alpha=(0,0,0)\trans$ and $\tau_{\mathrm{2b}}=(0,0.35,0.30)\trans$.
For this primary-null mean structure, compatibility occurred at $r=0$, when the primary residual was uncorrelated with the residuals of both active secondary endpoints.

\subsection{Simulation 2c: Binary Primary Endpoint}

Simulation 2c used a binary primary endpoint and two Gaussian secondary endpoints.
The latent-response model in \cref{eqn:common-dgp} used $K(0.30)$,
$$\alpha=(-1.18,0,0)\trans, \qquad \tau_{\mathrm{2c}}=(0.41,0.35,0.30)\trans,$$
and a primary latent residual variance fixed at one.
The primary intercept and treatment coefficient were chosen so that, after integrating over the covariate distribution,
$$\Pr(Y_{i1}=1\mid A_i=0)=0.15, \qquad \Pr(Y_{i1}=1\mid A_i=1)=0.25\mathperiod$$
Thus, the primary-endpoint estimand was a risk difference of 0.10.

The compatible latent Gaussian reference covariance was
\begin{equation}
    \label{eqn:binary-reference}
    \Sigma_{\epsilon,0}^*=\begin{pmatrix}1.00&0.45&0.38\\0.45&0.81&0.32\\0.38&0.32&0.89\end{pmatrix}\mathperiod
\end{equation}
We applied the same perturbation and grid as in Simulations 2a and 2b to the latent primary--secondary covariance vector.
Accordingly, $r=R^{2,*}_{1\mid-1,X}$ measured residual secondary-endpoint information on the latent Gaussian scale.
The covariance and treatment-effect structures were compatible at $r=0.30$.

\subsection{Simulation 3: Targeted Covariance Path Under the Global Null}

Simulation 3 used three Gaussian endpoints, $K(0.30)$, $\alpha=(0,0,0)\trans$, and $\tau_{\mathrm{3}}=(0,0,0)\trans$.
We began with the Gaussian reference covariance in \cref{eqn:gaussian-reference} and held every entry fixed except the covariance between the primary endpoint and the first secondary endpoint.
For a selected residual correlation $\rho_{12}$, we set
\begin{equation}
    \label{eqn:rho-path}
    \sigma_{12}(\rho_{12})=\rho_{12}\sqrt{\sigma_1^2\sigma_2^2}
\end{equation}
and replaced the $(1,2)$ and $(2,1)$ entries of the reference covariance by this value.
We used
$$\rho_{12}\in\{0,0.05,0.10,0.15,0.20,0.25,0.30,0.40,0.5227\}\mathperiod$$
The value 0.5227 is the residual correlation in the compatible Gaussian reference covariance.
Because $\sigma_{13}$ and the secondary covariance matrix $S$ remained fixed, $R^2_{1\mid-1,X}$ changed nonmonotonically as a derived feature of this path rather than serving as its directly controlled coordinate.

Under the global null, the SEM mean structure can be represented by setting the factor effect of treatment to zero.
Compatibility therefore depends on whether the residual covariance admits a one-factor decomposition
$$\Sigma_\epsilon=\lambda\lambda\trans+\operatorname{diag}(\theta), \qquad \theta_p\geq0,$$
rather than on alignment between nonzero treatment effects and factor loadings.
Among the simulated grid points, the residual covariance was factor-compatible at $\rho_{12}=0.25$, 0.30, 0.40, and 0.5227 and was incompatible at values from 0 through 0.20.

\section{Supplemental Simulation Results}
\label{sec:supp-sim-results}

\begin{figure}[!htbp]
    \centering
    \includegraphics[width=\linewidth]{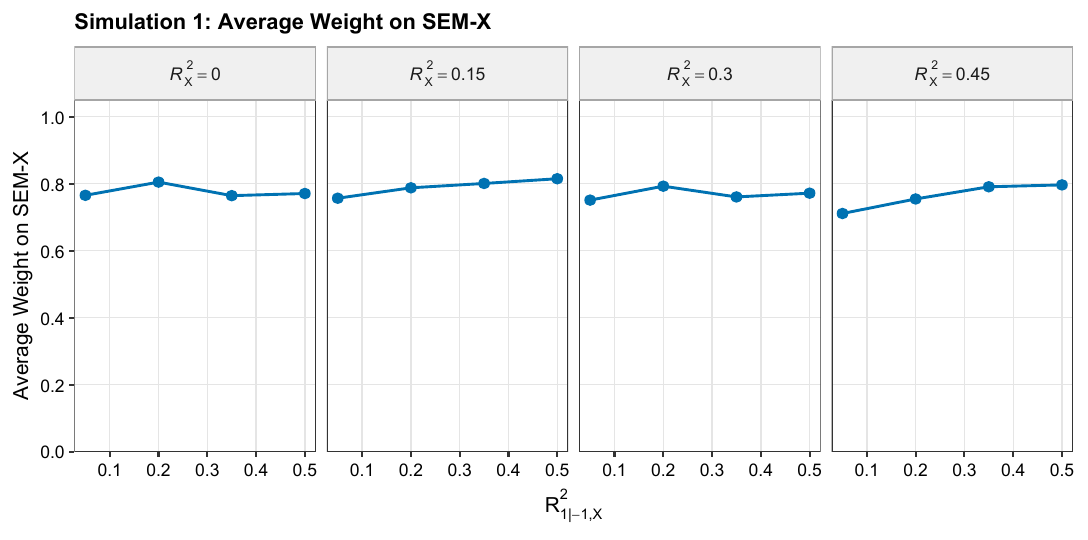}
    \caption{Average model-averaging weight assigned to $\widehat\tau_{1,\semx}$ in Simulation 1 as a function of residual secondary-endpoint information, with panels corresponding to the proportion of primary-endpoint variation explained by the baseline covariates.}
    \label{fig:sim1-average-weight}
\end{figure}

\begin{figure}
    \centering
    \includegraphics[width=\linewidth]{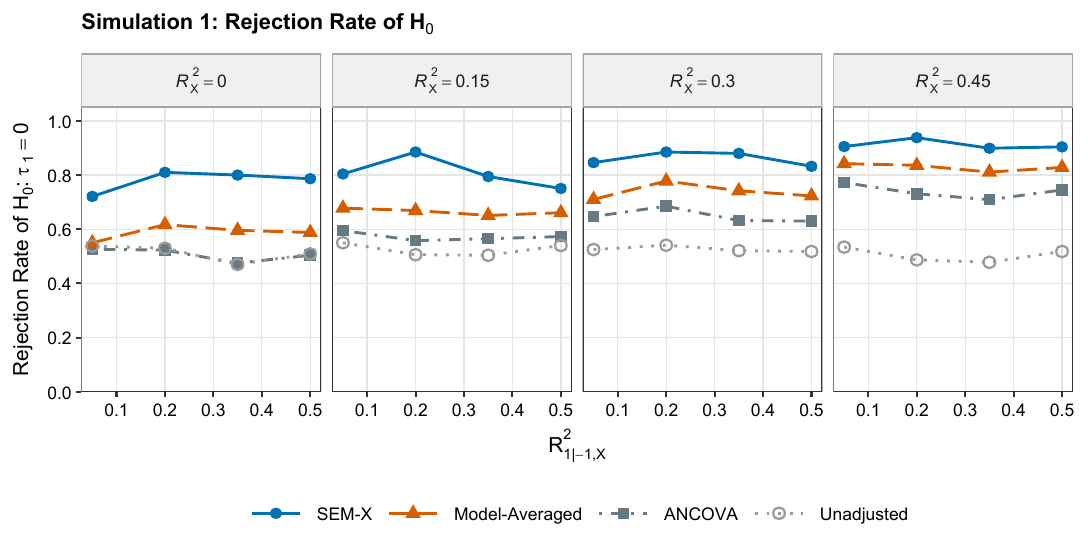}
    \caption{Rejection rate of the null hypothesis that $\tau_1=0$ for $\widehat\tau_{1,\semx}$, $\widehat\tau_{1,\mathrm{MA}}$, $\widehat\tau_{1,\mathrm{Adj}}$, and $\widehat\tau_{1,\mathrm{DM}}$ in Simulation 1. The treatment effect was nonzero in all scenarios, so the rejection rate represents power.}
    \label{fig:sim1-rejection}
\end{figure}

\begin{figure}
    \centering
    \includegraphics[width=\linewidth]{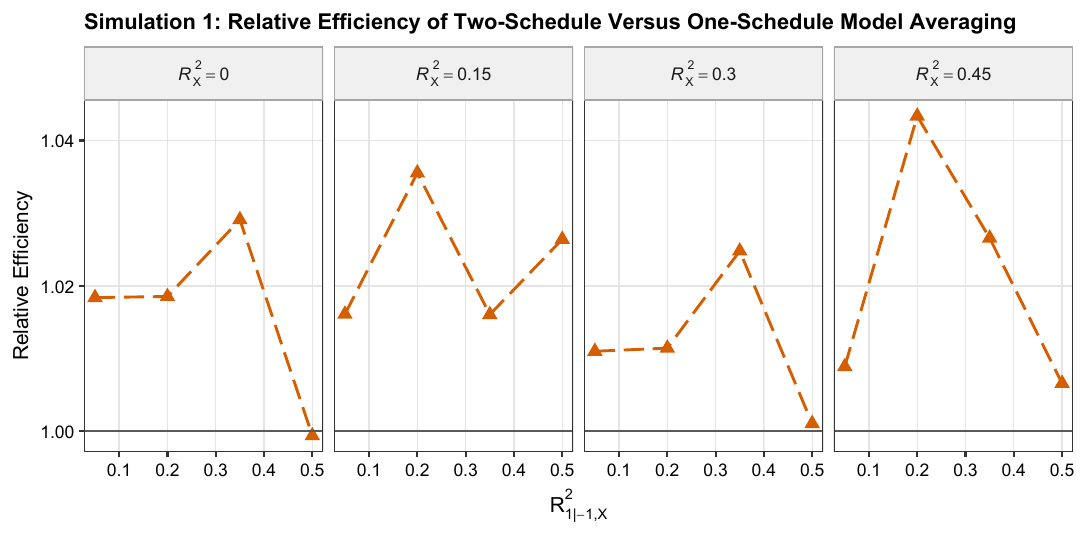}
    \caption{Relative efficiency of the model-averaged estimator using two versus one cross-validation schedule in Simulation 1. Values greater than one favor aggregation over two schedules; the horizontal reference line denotes equal efficiency.}
    \label{fig:sim1-cv-efficiency}
\end{figure}

\begin{figure}
    \centering
    \includegraphics[width=\linewidth]{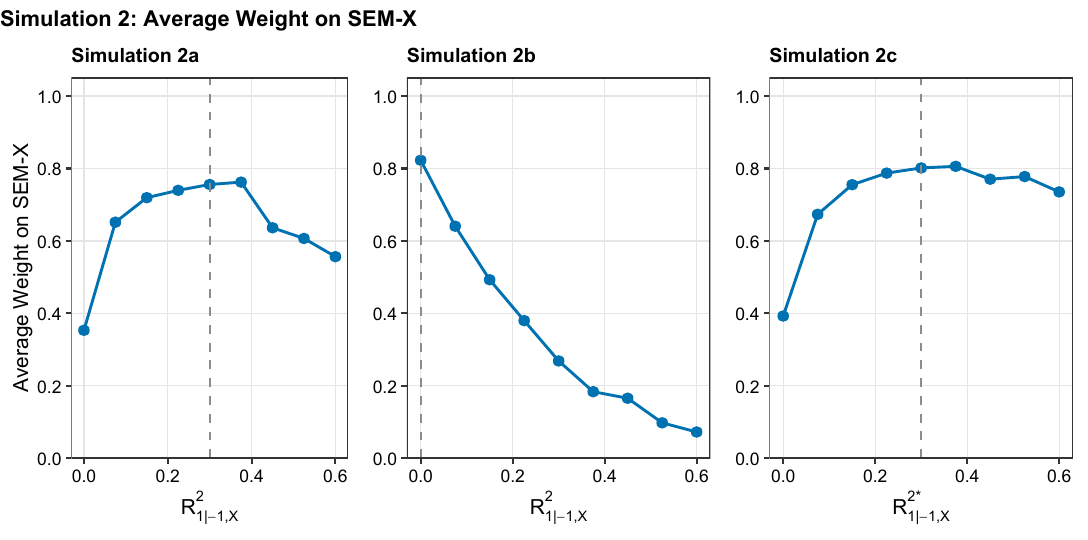}
    \caption{Average model-averaging weight assigned to $\widehat\tau_{1,\semx}$ in Simulations 2a--2c as a function of residual secondary-endpoint information. The quantity is defined on the latent Gaussian scale in Simulation 2c. Dashed vertical reference lines denote the covariance compatible with the structural equation model.}
    \label{fig:sim2-average-weight}
\end{figure}

\begin{figure}
    \centering
    \includegraphics[width=\linewidth]{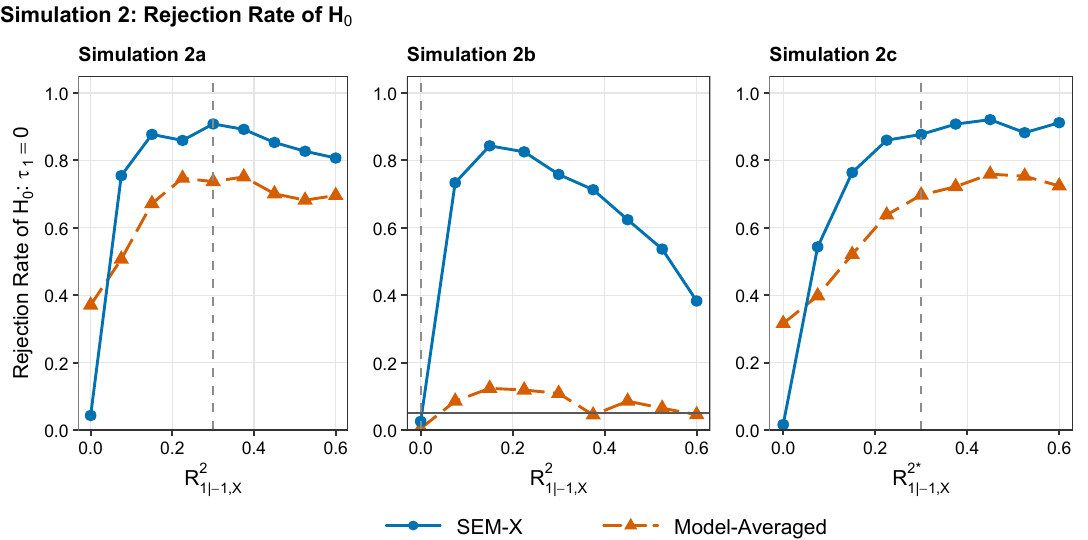}
    \caption{Rejection rate of the null hypothesis that $\tau_1=0$ for $\widehat\tau_{1,\semx}$ and $\widehat\tau_{1,\mathrm{MA}}$ in Simulations 2a--2c. Rejection rates represent power in Simulations 2a and 2c and type I error in Simulation 2b. Dashed vertical reference lines denote the covariance compatible with the structural equation model; the horizontal reference line in Simulation 2b denotes the nominal 0.05 level.}
    \label{fig:sim2-rejection}
\end{figure}

\begin{figure}
    \centering
    \includegraphics[width=\linewidth]{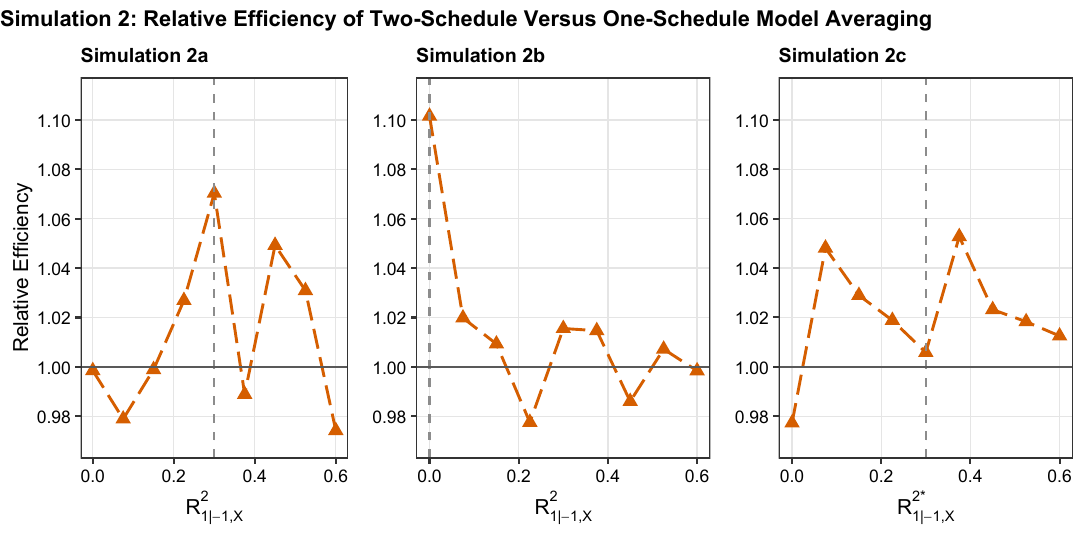}
    \caption{Relative efficiency of the model-averaged estimator using two versus one cross-validation schedule in Simulations 2a--2c. Values greater than one favor aggregation over two schedules; horizontal reference lines denote equal efficiency, and dashed vertical reference lines denote the covariance compatible with the structural equation model.}
    \label{fig:sim2-cv-efficiency}
\end{figure}

\begin{figure}
    \centering
    \includegraphics[width=\linewidth]{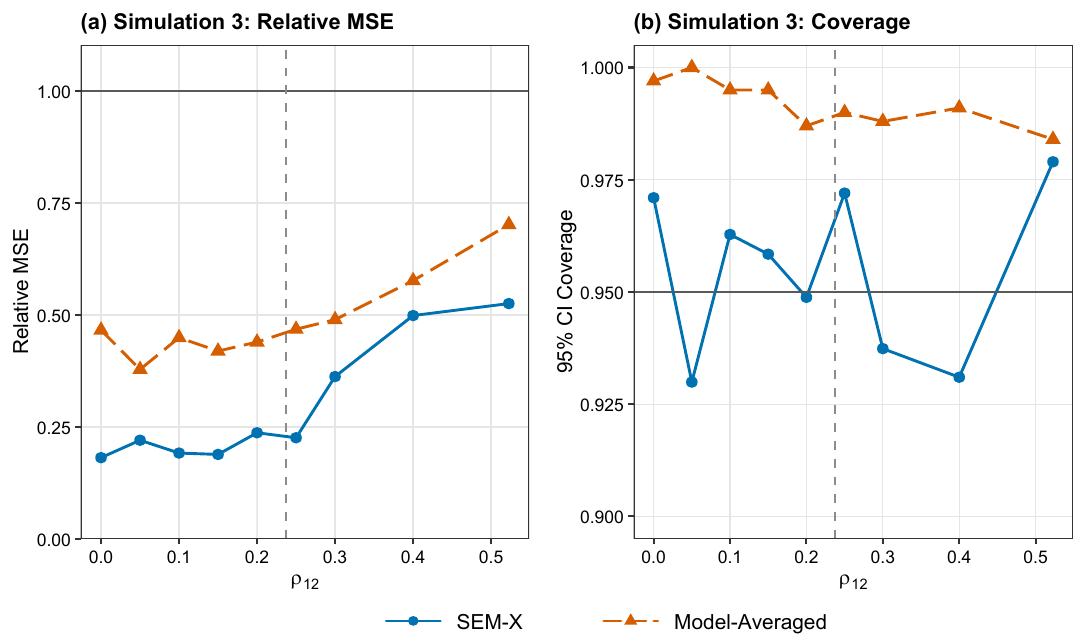}
    \caption{Relative mean squared error (MSE) versus $\widehat\tau_{1,\mathrm{Adj}}$ and coverage of 95\% confidence intervals for $\widehat\tau_{1,\semx}$ and $\widehat\tau_{1,\mathrm{MA}}$ under the global null in Simulation 3. The horizontal reference lines denote equal MSE and nominal 0.95 coverage, respectively; the dashed vertical line denotes the lower boundary of the factor-compatible covariance region.}
    \label{fig:sim3-mse-coverage}
\end{figure}

\begin{figure}
    \centering
    \includegraphics[width=\linewidth]{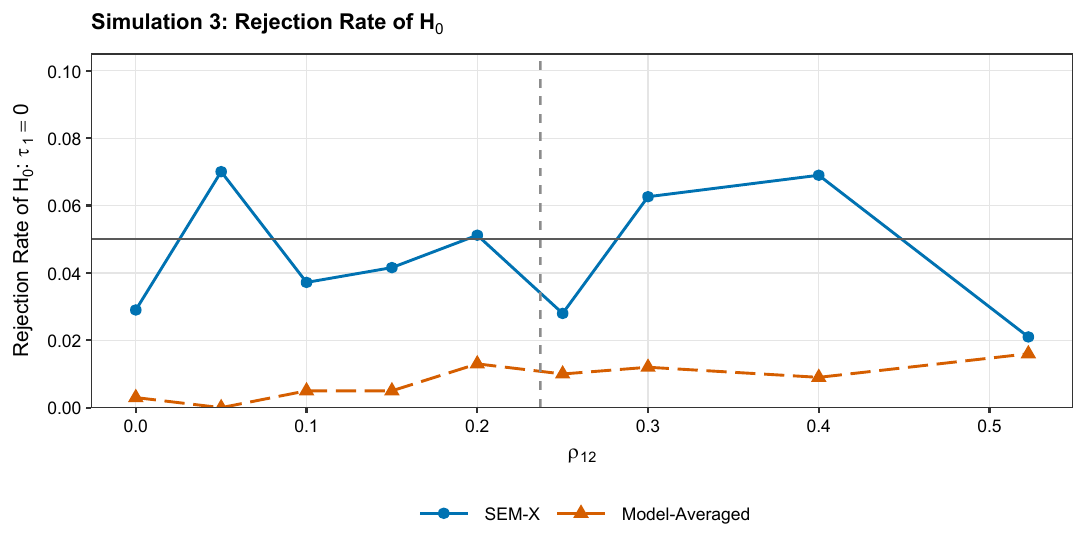}
    \caption{Rejection rate of the null hypothesis that $\tau_1=0$ for $\widehat\tau_{1,\semx}$ and $\widehat\tau_{1,\mathrm{MA}}$ under the global null in Simulation 3. The horizontal reference line denotes the nominal 0.05 level, and the dashed vertical line denotes the lower boundary of the factor-compatible covariance region.}
    \label{fig:sim3-rejection}
\end{figure}

\begin{figure}
    \centering
    \includegraphics[width=\linewidth]{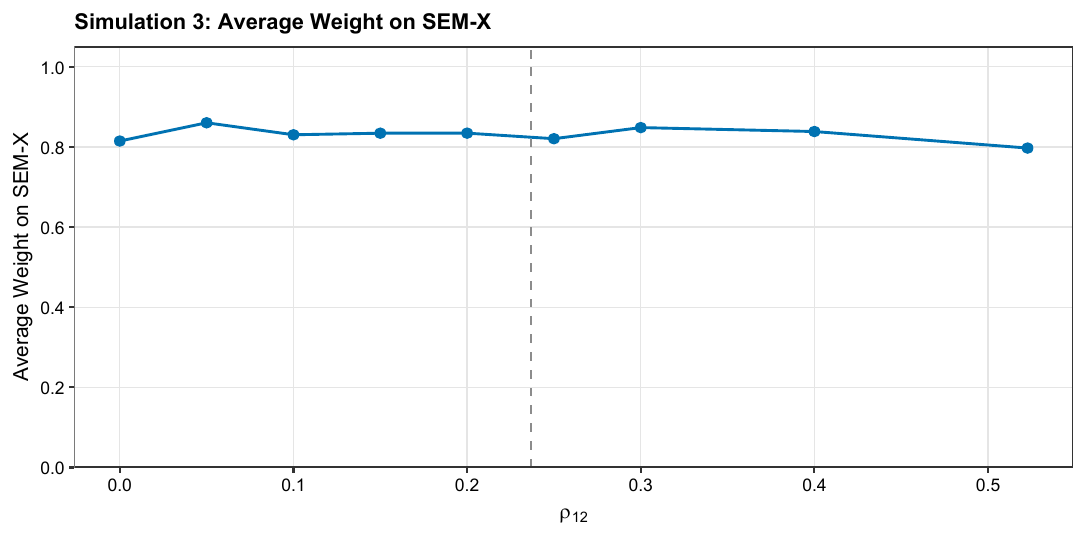}
    \caption{Average model-averaging weight assigned to $\widehat\tau_{1,\semx}$ under the global null in Simulation 3. The dashed vertical line denotes the lower boundary of the factor-compatible covariance region.}
    \label{fig:sim3-average-weight}
\end{figure}

\begin{figure}
    \centering
    \includegraphics[width=\linewidth]{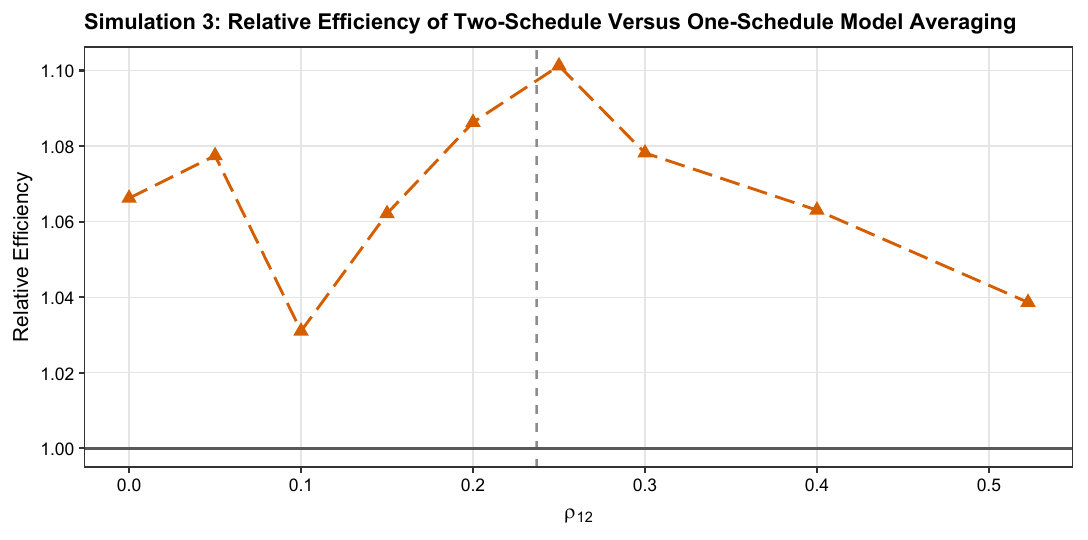}
    \caption{Relative efficiency of the model-averaged estimator using two versus one cross-validation schedule in Simulation 3. Values greater than one favor aggregation over two schedules; the horizontal reference line denotes equal efficiency, and the dashed vertical line denotes the lower boundary of the factor-compatible covariance region.}
    \label{fig:sim3-cv-efficiency}
\end{figure}

\FloatBarrier
\endgroup

\end{document}